\documentclass[reprint, twocolumn, showkeys]{revtex4}

\usepackage{graphicx}% Include figure files
\usepackage{dcolumn}% Align table columns on decimal point
\usepackage{bm}% bold math
\usepackage{color}% bold math
\usepackage{amsmath,amsthm}
\usepackage{amssymb,bm}
\usepackage{mathrsfs}
\usepackage{comment}
\usepackage{ulem}

\begin{document}

\title{Direct and ultrafast probing of quantum many-body interaction and Mott-insulator transition through coherent two-dimensional spectroscopy}
\author{Nguyen Thanh Phuc}
%\email{nthanhphuc@ims.ac.jp}
\email{nthanhphuc@moleng.kyoto-u.ac.jp}
%\affiliation{Department of Theoretical and Computational Molecular Science, Institute for Molecular Science, Okazaki 444-8585, Japan}
%\affiliation{Department of Structural Molecular Science, The Graduate University for Advanced Studies, Okazaki 444-8585, Japan}
\affiliation{Department of Molecular Engineering, Graduate School of Engineering, Kyoto University, Kyoto 615-8510, Japan}
\author{Pham Quang Trung}
\affiliation{Section of Brain Function Information, Supportive Center for Brain Research, National Institute for Physiological Sciences, Okazaki 444-8585, Japan}
%\author{Akihito Ishizaki}
%\affiliation{Department of Theoretical and Computational Molecular Science, Institute for Molecular Science, Okazaki 444-8585, Japan}
%\affiliation{Department of Structural Molecular Science, The Graduate University for Advanced Studies, Okazaki 444-8585, Japan}

%%%%%%%%%%%%%%%%%%%%%%%%%%
\begin{abstract}
Interactions between particles in quantum many-body systems play a crucial role in determining the electric, magnetic, optical, and thermal properties of the system. The recent progress in the laser-pulse technique has enabled the manipulations and measurements of physical properties on ultrafast timescales. Here, we propose a method for the direct and ultrafast probing of quantum many-body interaction through coherent two-dimensional (2D) spectroscopy. Up to a moderate interaction strength, the inter-particle interaction manifests itself in the emergence of off-diagonal peaks in the 2D spectrum before all the peaks coalesce into a single diagonal peak as the system approaches the Mott-insulating phase in the strongly interacting regime. The evolution of the 2D spectrum as a function of the time delay between the second and third laser pulses can provide important information on the ultrafast time variation of the interaction. 
\end{abstract}


\maketitle

%%%%%%%%%%%%%%%%%%%%%
\textit{Introduction--}
Important physical systems and materials often consist of a macroscopically large number of atoms, molecules, and electrons.
The interaction between constituent particles can dictate various physical properties of the system, including, for example, the electrical and thermal transports and the magnetic and optical properties.
However, in some metals and semiconductors, owing to the screening of the Coulomb interaction between electrons, their low-energy behaviors are similar to those of a system of non-interacting particles. These systems can be described by Landau's Fermi liquid theory~\cite{Landau56}. 
In contrast, there also exist strongly correlated electronic systems beyond the Fermi liquid theory, in which the interaction between electrons cannot be ignored. Examples include the Luttinger liquid in one dimension~\cite{Tomonaga50, Luttinger63}, heavy fermionic systems~\cite{Stewart01}, and systems near critical points such as the Mott-insulator phase transition~\cite{Mott68, Imada98}. 
The interparticle interaction can also result in exotic phenomena such as superfluidity and superconductivity~\cite{Leggett-book}. 
%Strong interparticle interactions and their effects on the static and dynamic properties of the system have also been investigated extensively in ultracold atomic and molecular systems~\cite{Bloch08, Bloch12, Gross17, Moses17, Bohn17}, where the ratio of the interaction energy to the kinetic energy can be varied under control. For example, the absorption spectroscopic measurements of the superfluid-to-Mott-insulator phase transition were performed with rubidium and ytterbium atoms~\cite{Campbell06, Kato16}.

Meanwhile, continual progress in the development of laser-pulse techniques has enabled a faster manipulation and measurement of the physical and chemical properties of electronic, atomic, and molecular systems~\cite{Krausz09}. 
This allowed us to observe intriguing nonequilibrium phenomena, such as light-induced superconductivity~\cite{Fausti11, Cavalleri18}, ultrafast spintronics~\cite{Kimel05, Kirilyuk10}, and the Floquet engineering of electronic band topology~\cite{Wang13, McIver20}.
In this Letter, we propose a method to investigate the interaction between particles directly in quantum many-body systems on an ultrafast timescale using coherent two-dimensional (2D) spectroscopy. 
Coherent multi-dimensional, especially 2D, spectroscopy has been widely used to study electronic excitation (exciton) and vibration dynamics in molecular and semiconductor systems~\cite{Mukamel-book, Mukamel00, Jonas03, Cho08, Cho-book, Abramavicius09, Hamm-book, Cundiff09, Cohen11}. More recently, intersubband electronic excitations in quantum wells~\cite{Kuehn11}, carrier dynamics in graphene~\cite{Woerner13}, spin-wave~\cite{Lu17} and fractional excitations~\cite{Wan19, Choi20} in magnetic materials, marginal Fermi glass~\cite{Mahmood20}, and high-temperature superconductors~\cite{Novelli20} have been studied. 
%In coherent 2D spectroscopy, a sequence of three laser pulses is used to excite the system, and the subsequent coherent light emission induced by the polarization of the system is measured. 
%The 2D spectrum displays the emitted optical signal as a function of the frequencies $\omega_1$ and $\omega_3$, which is obtained by performing a Fourier transformation with respect to the time interval $t_1$ between the first two pulses and the time interval $t_3$ between the third pulse and the emitted signal, respectively.
%Physically, $\omega_1$ and $\omega_3$ amount to the excitation and emission frequencies, respectively.
%The diagonal/off-diagonal peaks in the coherent 2D spectrum represent processes with equal/unequal excitation and emission frequencies. 
%Importantly, off-diagonal peaks can emerge only if the two transitions associated with the optical excitation and emission are coupled to each other~\cite{Dai10, Khalil03, Zheng05}.
%However, notably, the coupling between two transitions is purely at the level of single-body physics, whereas the much more complex quantum many-body physics is investigated in this study.

\begin{figure}[t] % float placement: (h)ere, page (t)op, page (b)ottom, other (p)age
  \centering
  % file name: E:/Draft-Coherent 2D spectroscopy of interacting quantum many-body systems (Feb 2020)/Fig1.eps
  \includegraphics[width=3.4in, keepaspectratio]{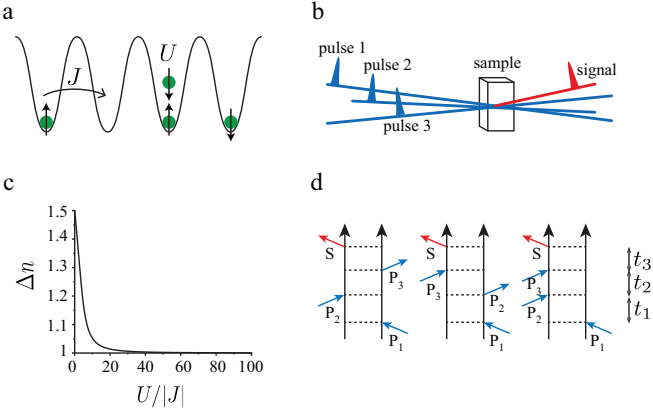}
  \caption{Coherent 2D spectroscopic measurement of an interacting quantum many-body system. (a) System of spin-1/2 fermionic particles moving in a lattice. The two parameters of the Hubbard model are the hopping amplitude $J$ between neighboring sites and the on-site interaction $U$ between two particles at the same site. (b) Setup of the coherent 2D spectroscopic measurement. Three laser pulses successively interact with the system, and the light-emission signal induced by the polarization of the system is measured. (c) Variance $\Delta n$ of the number of particles per site in the quantum many-body ground state of the system. Here, the number of lattice sites is $N=6$, the filling factor is one-half, i.e., the number of particles is equal to the number of sites, the total magnetization is zero, and the periodic boundary condition is applied. (d) Double-sided Feynman diagrams for the light-matter interaction processes involved in the rephasing (photon echo) signal.}
  \label{fig: system setup}
\end{figure}

Here, we calculate the coherent 2D spectrum of an interacting quantum many-body system of spin-1/2 fermions (see Figs.~\ref{fig: system setup}a,b), which can be, for example, electrons moving in a crystal lattice or ultracold atoms/molecules moving in an artificial lattice. We observe that a system of non-interacting particles would display a 2D spectrum with only peaks lying on the diagonal axis. 
This is attributed to the fact that the quasimomentum of the particles is a good quantum number in the absence of interparticle interaction. Moreover, because of the conservation of momentum in the light-matter interaction, two optical transitions with different quasi-momenta are not coupled to each other.
In contrast, if the interaction between particles is sufficiently strong, off-diagonal peaks emerge in the 2D spectrum. Thus, this can be used as a direct probe of a nonnegligible interparticle interaction. 
In an interacting quantum many-body system, quasimomentum states are no longer energy eigenstates of the system. Consequently, two transitions with different optical excitation and emission frequencies can be effectively coupled to each other, leading to the emergence of off-diagonal peaks.
However, if the interaction strength is increased further, when the system approaches the Mott-insulating phase, in which exactly one particle is localized at each lattice site as the hopping of particles between neighboring sites becomes energetically unfavorable, all the peaks in the 2D spectrum coalesce into a single diagonal peak at the frequency of the excitation band.
Consequently, coherent 2D spectroscopy can be used to investigate the entire range of interaction strength from weak to moderate and strong interaction regimes.
We also investigate the coherent 2D spectrum of the system when the interaction energy relative to the kinetic energy varies with time. 
The evolution of the 2D spectrum as a function of the time delay $t_2$ between the second and third pulses can provide us with important information on the ultrafast time variation of the interaction. 
%Notably, in contrast to the other schemes~\cite{Gessner14, Schlawin14}, our proposed method to probe the quantum many-body interaction does not require single-site addressability, which might be feasible for ultracold atoms and ions but is difficult for electronic systems.

%%%%%%%%%%%%%%%%%%
\textit{System--} 
Consider the Hubbard model of a system of spin-1/2 fermionic particles moving in a lattice, whose Hamiltonian for the ground-state band is given by
\begin{align}
\hat{H}_\mathrm{g}=&
-J\sum_{\langle i,j\rangle} \sum_{\sigma=\uparrow,\downarrow} \left( \hat{c}_{i,\sigma}^\dagger \hat{c}_{j,\sigma} + \text{h.c.}\right) +U\sum_j \hat{n}_{j,\uparrow}\hat{n}_{j, \downarrow},
\end{align} 
where $\hat{c}_{j, \sigma}$ denotes the annihilation operator of a particle with spin $\sigma$ located at the $j$th site, and $\hat{n}_{j,\sigma}=\hat{c}_{j, \sigma}^\dagger \hat{c}_{j, \sigma}$ is the particle number operator. The parameters $J$ and $U$ represent the hopping amplitude of the particle between neighboring sites and the on-site interaction between two particles located at the same site, respectively. Here, $\langle ...\rangle$ denotes a pair of nearest neighboring sites, and h.c. stands for Hermitian conjugate. Here, we restrict our consideration to the case of the half-filling factor, i.e., the number of particles is equal to the number of lattice sites. Furthermore, the total magnetization of the system is zero, i.e., the number of particles with spin-up is equal to that of particles with spin-down. 

For a weakly interacting system with $|J|\gg U$, the motions of particles are almost independent of one another (except for the Pauli exclusive principle) and their wavefunctions are delocalized over different sites in the lattice. In contrast, in the strongly interacting limit $|J|\ll U$, the system would be in the Mott insulating phase, where each lattice site is occupied by exactly one particle as the hopping of particles between neighboring sites is energetically unfavorable.
This transition is reflected by the change in the variance $\Delta n=\langle \psi_0| \hat{n}_j^2|\psi_0 \rangle$ of the number of particles per site for the quantum many-body ground state $|\psi_0\rangle$. This is shown in Fig.~\ref{fig: system setup}c as a function of $U/|J|$ for a system of $N=6$ particles in a one-dimensional lattice. Here, $\hat{n}_j=\hat{n}_{j,\uparrow}+\hat{n}_{j,\downarrow}$. 
The variance is maximum for $U=0$, at which $\Delta n=1.5$; it decreases with increasing interaction strength and approaches $\Delta n=1$ for a sufficiently strong interaction. 

We consider a one-dimensional system with the periodic boundary condition. As for the optical transitions caused by the light-matter interactions with the laser pulses in coherent 2D spectroscopy, we consider a Hubbard model with two energy bands: the ground-state and first excited bands. The total Hamiltonian of the system is given by
\begin{align}
\hat{H}=&
-\sum_{j=0}^{N-1}\sum_{\sigma=\uparrow,\downarrow} 
\left(J_\mathrm{g} \hat{c}_{\mathrm{g}, j+1,\sigma}^\dagger \hat{c}_{\mathrm{g}, j,\sigma} 
+J_\mathrm{e} \hat{c}_{\mathrm{e}, j+1,\sigma}^\dagger \hat{c}_{\mathrm{e}, j,\sigma} + \text{h.c.}\right) \nonumber\\
&+\sum_{j=0}^{N-1}\sum_{\alpha\not=\beta}
U_{\alpha\beta} \hat{n}_{j,\alpha}\hat{n}_{j, \beta}
+\epsilon_\mathrm{eg}\sum_{j=0}^{N-1}\sum_{\sigma=\uparrow,\downarrow}
\hat{c}_{\mathrm{e}, j,\sigma}^\dagger \hat{c}_{\mathrm{e}, j,\sigma},
\label{eq: quantum many-body Hamiltonian}
\end{align} 
where $\alpha, \beta=(\mathrm{g/e}; \uparrow/\downarrow)$. Here, $J_\mathrm{g}$ and $J_\mathrm{e}$ denote the hopping amplitudes of particles between neighboring sites in the ground and excited bands, respectively. Owing to the difference in the spatial confinement of the wavefunction of the particle at the lattice sites between the two bands, $J_\mathrm{g}\not=J_\mathrm{e}$. In general, the on-site interaction should also depend on the spins of the particles and on whether they are in the ground or excited band; however, for simplicity, we assume that the on-site interaction is characterized by a single variable parameter, namely $U_{\alpha\beta}=U$. By shifting the origin of the 2D spectrum by an amount equal to the excitation energy $\epsilon_\mathrm{eg}$, we can set $\epsilon_\mathrm{eg}=0$ without loss of generality. The periodic boundary condition indicates that $j=N$ is equivalent to $j=0$. 

The optical signal measured via coherent 2D spectroscopy is proportional to $i P^{(3)}(t)$, where $P^{(3)}(t)$ is the time-dependent third-order polarization of the system. It can be expressed as a convolution of the third-order response function $R^{(3)}(t_1, t_2, t_3)$ and the electric fields of the lasers~\cite{Mukamel-book, Hamm-book}
\begin{align}
P^{(3)}(t)=&
\int_0^\infty \text{d}t_1 \int_0^\infty \text{d}t_2 \int_0^\infty \text{d}t_3 R^{(3)}(t_1, t_2, t_3) \nonumber\\
& E(t-t_3) E(t-t_3-t_2) E(t-t_3-t_2-t_1).
\end{align}
As the wavelength of the laser is typically much larger than the lattice constant of the crystal lattice for electrons, the electric field can be considered homogeneous over a large number of lattice sites. A similar nanoscale artificial lattice for ultracold atoms can be realized by using nanoplasmonic systems~\cite{Gullans12}, photonic crystals~\cite{Tudela15}, time-periodic modulations~\cite{Nascimbene15}, and superconductors~\cite{Isart13}.

In the impulsive limit of the laser pulses, where the electric field is given by a sum of three Dirac's delta functions, the polarization is proportional to the nonlinear response function $R^{(3)}(t_1, t_2, t_3)$ , where $t_1$, $t_2$, and $t_3$ are the time intervals between the laser pulses and the emitted signal (see Fig.~\ref{fig: system setup}). 
The nonlinear response function and emitted signal are generated by various processes, each of which involves four interactions between light and matter. 
These processes can be grouped into three categories according to the direction of the emitted signal: rephasing, non-rephasing, and double quantum coherence. 
The three types of signals can generally provide different types of information about the energy levels and dynamics of the system. In this study, either rephasing or non-rephasing signals can be used as we concentrate on the interaction-induced emergence of off-diagonal peaks.
The rephasing or photon-echo signal is detected in the direction given by the vector $\mathbf{k}_\mathrm{r}=-\mathbf{k}_1+ \mathbf{k}_2+\mathbf{k}_3$. 
The light-matter interactions for the processes included in the rephasing signal are illustrated by the double-sided Feynman diagram (Fig.~\ref{fig: system setup}d). The corresponding rephasing third-order response function $R_\mathrm{r}^{(3)}(t_1, t_2, t_3)$ can be expressed in terms of the Liouville-space operators as~\cite{Mukamel-book}
\begin{align}
&R_\mathrm{r}^{(3)}(t_1, t_2, t_3) \nonumber\\
=& \left( \frac{i}{\hbar}\right)^3
\text{Tr}\left\{\hat{\mu}_\leftarrow \mathcal{G}(t_3) \hat{\mu}^\times_\rightarrow \mathcal{G}(t_2) \hat{\mu}^\times_\rightarrow \mathcal{G}(t_1) \hat{\mu}^\times_\leftarrow \hat{\rho}_0 \right\}, 
\end{align}
where $\hat{\rho}_0=|\psi_0\rangle \langle \psi_0|$ is the density operator for the quantum many-body ground state of the system, and the transition dipole moment operators $\hat{\mu}_{\leftarrow/\rightarrow}$ are given by
\begin{align}
\hat{\mu}_\leftarrow=&\sum_{j=0}^{N-1}\sum_{\sigma=\uparrow,\downarrow} \mu_\mathrm{ge}\hat{c}_{\mathrm{g}, j, \sigma}^\dagger \hat{c}_{\mathrm{e}, j, \sigma}, \\
\hat{\mu}_\rightarrow=&\sum_{j=0}^{N-1}\sum_{\sigma=\uparrow,\downarrow} \mu_\mathrm{eg}\hat{c}_{\mathrm{e}, j, \sigma}^\dagger \hat{c}_{\mathrm{g}, j, \sigma}.
\end{align}
Here, $\mu_\mathrm{eg}$ is the transition dipole moment between the ground and excited bands, and $\mu_\mathrm{ge}=\mu_\mathrm{eg}^*$. The superoperators in Liouville space are defined as $\hat{\mu}^\times \hat{\rho}=\hat{\mu}\hat{\rho}-\hat{\rho}\hat{\mu}$ and $\mathcal{G}(t)\hat{\rho}=e^{-i\hat{H}t/\hbar}\hat{\rho}e^{i\hat{H}t/\hbar}$, where $\hat{H}$ is the quantum many-body Hamiltonian of the system given in Eq.~\eqref{eq: quantum many-body Hamiltonian}. 
A small dephasing rate $\kappa$ is introduced to account for the dephasing of the quantum coherence between the ground and an excited states during the time intervals $t_1$ and $t_3$.
The coherent 2D rephasing spectrum is obtained by making a Fourier transformation of the emitted signal $S(t_1, t_2, t_3)$ with respect to the time intervals $t_1$ and $t_3$,
\begin{align}
S(\omega_1, t_2, \omega_3)=&
\int_0^\infty \text{d}t_1  \int_0^\infty \text{d}t_3 e^{i(\omega_1t_1+\omega_3t_3)} S(t_1, t_2, t_3).
\end{align}
In the following 2D spectra, the real part of $S(\omega_1, t_2, \omega_3)$ is plotted as a function of $-\omega_1$ and $\omega_3$. 

\begin{figure}[tbp] % float placement: (h)ere, page (t)op, page (b)ottom, other (p)age
  \centering
  % file name: E:/Draft-Coherent 2D spectroscopy of interacting quantum many-body systems (Feb 2020)/Fig2.eps
  \includegraphics[keepaspectratio]{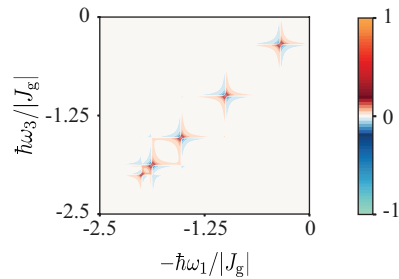}
  \caption{Coherent 2D rephasing spectrum of a system of noninteracting spin-1/2 fermionic particles. The number of lattice sites is $N=18$. The spectral intensity is normalized by its maximum value and represented by the color scale. The frequencies $\omega_1$ and $\omega_3$ are normalized by the hopping amplitude $|J_\mathrm{g}|$ between neighboring sites of particles moving in the ground band. Here, the origin of the 2D spectrum was shifted by the excitation energy $\epsilon_\mathrm{eg}$ between the ground and excited bands.}
  \label{fig: 2D spectrum for non-interacting systems}
\end{figure}

We first consider a system of noninteracting particles, namely $U=0$. The coherent 2D rephasing spectrum for $N=18$ is shown in Fig.~\ref{fig: 2D spectrum for non-interacting systems}. Here, the system's parameters were considered as follows: $J_\mathrm{e}/J_\mathrm{g}=2$ and $\hbar\kappa/|J_\mathrm{g}|=0.01$. The time delay was set to $t_2=0$.
It is evident that all the peaks lie on the diagonal axis $\omega_3=-\omega_1$. The number of peaks increases with the lattice size. In the thermodynamic limit ($N\to\infty$), the peaks are distributed on the segment $0\leq \hbar\omega_1\leq 2(J_\mathrm{e}-J_\mathrm{g})$ of the diagonal axis. 
Note that the origin of the 2D spectrum was shifted by the excitation energy between the ground and excited bands, namely $\tilde{\omega}_1=\omega_1+\epsilon_\mathrm{eg}$ and $\tilde{\omega}_3=\omega_3-\epsilon_\mathrm{eg}$. Thus, the origin $\tilde{\omega}_1=\tilde{\omega}_3=0$ of the 2D spectrum corresponds to $\omega_3=-\omega_1=\epsilon_\mathrm{eg}$. 

In the absence of interaction between particles, quasi-momentum is a good quantum number. The single-particle energy eigenstates in the ground and excited bands are characterized by a wavevector $k$ (within the Brillouin zone) with the corresponding energy eigenvalues given by $\epsilon_k^\mathrm{g}=-2J_\mathrm{g}\cos\left(2\pi k/N\right)$ and $\epsilon_k^\mathrm{e}=\epsilon_{eg}-2J_\mathrm{e}\cos\left(2\pi k/N\right)$ with $k=-N/2, \cdots, N/2$. 
Owing to the conservation of momentum in the light-matter interaction, optical transitions can only occur between pairs of single-particle energy eigenstates in the ground and excited bands with the same wavevector. 
As these transitions are not coupled to one another, only diagonal peaks appear in the coherent 2D rephasing spectrum. 
With the half-filling factor, the quantum many-body ground state of the system contains all the single-particle energy eigenstates of $\hat{H}_\mathrm{g}$ with the wavevectors $-N/4\leq k \leq N/4$ (for $J_\mathrm{g}>0$). 
The corresponding transition energy satisfies $\epsilon_{eg}-2(|J_\mathrm{e}|-|J_\mathrm{g}|)\leq \epsilon_k^\mathrm{e}-\epsilon_k^\mathrm{g}\leq\epsilon_{eg}$.

\begin{figure}[tbp] % float placement: (h)ere, page (t)op, page (b)ottom, other (p)age
  \centering
  % file name: E:/Draft-Coherent 2D spectroscopy of interacting quantum many-body systems (Feb 2020)/Fig3.eps
  \includegraphics[keepaspectratio]{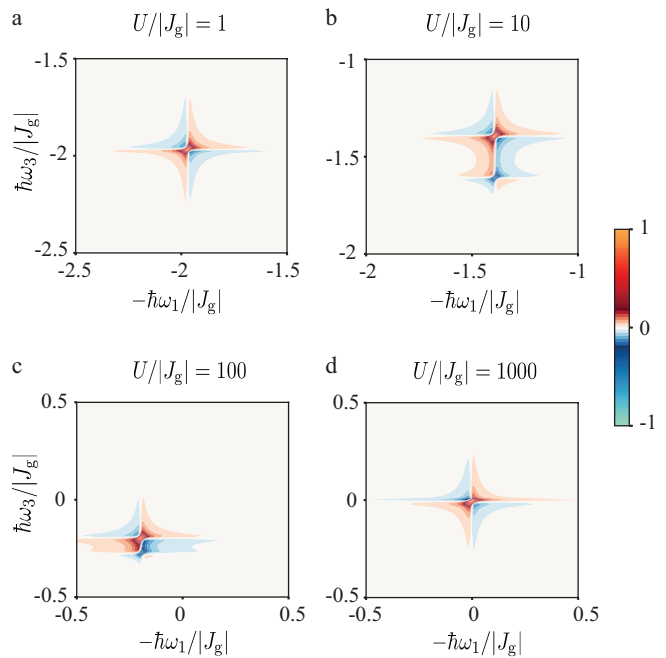}
  \caption{Coherent 2D rephasing spectra of a system of interacting spin-1/2 fermionic particles with different interaction strengths. Here, the interaction strength $U$ is normalized by the hopping amplitude $|J_\mathrm{g}|$ and the number of lattice sites is $N=2$.}
  \label{fig: 2D spectrum of interacting systems with different interaction strengths}
\end{figure}

Next, we consider a system of interacting particles, namely $U>0$ (repulsive interaction). As the computational cost increases exponentially with the system size, we restrict the computation to a small system of $N=2$ lattice sites. 
The coherent 2D rephasing spectra for different interaction strengths (normalized by the hopping amplitude $|J_\mathrm{g}|$) are shown in Fig.~\ref{fig: 2D spectrum of interacting systems with different interaction strengths}. 
The other parameters of the system are the same as in the case of noninteracting particles.  
As the interparticle interaction becomes stronger, an off-diagonal peak starts to emerge in addition to the diagonal peak. 
The off-diagonal peak persists up to an interaction of $U/|J_\mathrm{g}|\simeq 100$.
Finally, at the strongly interacting limit $U/|J_\mathrm{g}|\simeq 1000$, all the peaks coalesce into a single diagonal peak at the origin of the 2D spectrum. 
In the presence of the interparticle interaction, the quantum many-body ground state cannot be expressed by a collection of single-particle states. Consequently, the optical transitions become effectively coupled to one another, leading to the emergence of off-diagonal peaks.
In the strongly interacting limit, the quantum many-body ground state is in the Mott insulating phase with each lattice site being occupied by exactly one particle. 
The optical excitation of each particle occurs locally with an excitation energy equal to $\epsilon_\mathrm{eg}$. This results in a single diagonal peak at the origin of the 2D spectrum.

\begin{figure}[tbp] % float placement: (h)ere, page (t)op, page (b)ottom, other (p)age
  \centering
  % file name: E:/Draft-Coherent 2D spectroscopy of interacting quantum many-body systems (Feb 2020)/Fig4.eps
  \includegraphics[keepaspectratio]{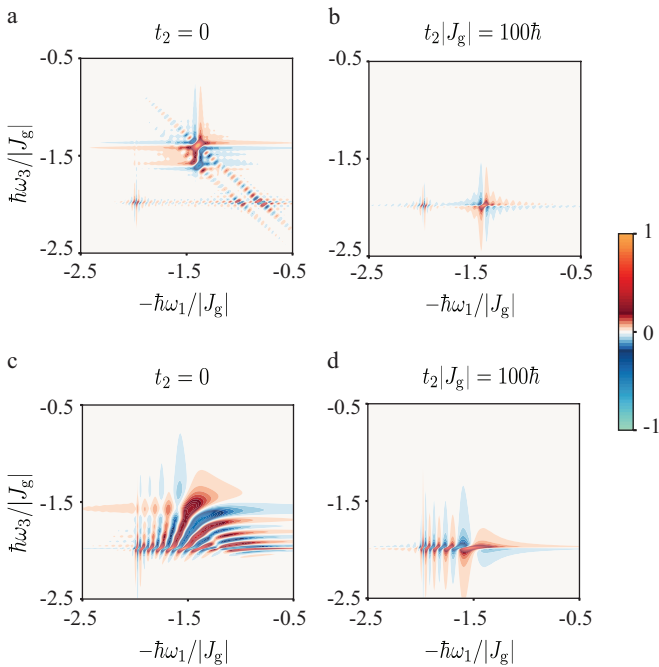}
  \caption{Coherent 2D rephasing spectra of a system of interacting spin-1/2 fermionic particles as a function of the time delay $t_2$ between the second and third laser pulses. Upper panels (a), (b): the interaction strength is abruptly switched off from $U/|J_\mathrm{g}|=10$ to $U=0$ at time $\tau|J_\mathrm{g}|=100\hbar$. Lower panels (c), (d): the interaction strength is switched off steadily over the time interval of $\tau$.}
  \label{fig: 2D spectrum for time-varying interaction strength}
\end{figure}

Finally, we consider the case of a time-varying interaction $U(t)$ and investigate the coherent 2D spectrum as a function of the time delay $t_2$. 
If the interaction is switched off abruptly at time $\tau$ after the incidence of the first laser pulse, i.e., $U(t)=U_0$ for $t\leq \tau$ and $U(t)=0$ for $t>\tau$, the coherent 2D spectra for different time delays are shown in Figs.~\ref{fig: 2D spectrum for time-varying interaction strength}a, b. 
Here, the initial interaction strength is $U_0/|J_\mathrm{g}|=10$, the switching time is $\tau|J_\mathrm{g}|=100\hbar$, and the other parameters of the system are the same as in the case of a time-independent interaction.
The 2D spectrum is observed to change from $t_2=0$ to $t_2=\tau$, and subsequently, it remains almost unchanged. 
It can be observed that, at $t_2=0$, the pair of a diagonal peak at $\hbar\omega_1=-\hbar\omega_3\simeq 1.4|J_\mathrm{g}|$ and an off-diagonal peak at $\hbar\omega_1\simeq 1.4|J_\mathrm{g}|, \hbar\omega_3\simeq -1.6|J_\mathrm{g}|$, which is the 2D spectrum for constant $U/|J_\mathrm{g}|=10$, is extended along the diagonal direction of $\omega_1=\omega_3$ to $\hbar\omega_3= -2|J_\mathrm{g}|$. 
The spectrum also contains a small diagonal peak at $\hbar\omega_1=-\hbar\omega_3=2|J_\mathrm{g}|$, which is the 2D spectrum for a constant $U=0$.
The 2D spectrum changes with the variable time delay. At $t_2=\tau$, it is composed of a small diagonal peak at $\hbar\omega_1=-\hbar\omega_3=2|J_\mathrm{g}|$, an off-diagonal peak at $\hbar\omega_1\simeq 1.4|J_\mathrm{g}|, \hbar\omega_3=-2|J_\mathrm{g}|$, and a connection between them. 

If the interaction is switched off steadily over a time period of $\tau$ after the incidence of the first laser pulse, i.e., $U(t)=U_0(1-t/\tau)$ for $t\leq \tau$ and $U(t)=0$ for $t>\tau$, the coherent 2D spectra for different time delays are shown in Figs.~\ref{fig: 2D spectrum for time-varying interaction strength}c, d.
A fringe pattern emerges in the 2D spectrum, which should be attributed to the continuous time variation of the interaction strength.
The spectrum changes gradually with a variable time delay. At $t_2=\tau$, it consists of an array of peaks distributed on the segment between $\hbar\omega_1\simeq 1.4|J_\mathrm{g}|$ and $\hbar\omega_1=2|J_\mathrm{g}|$ of the horizontal line $\hbar\omega_3=-2|J_\mathrm{g}|$.

%%%%%%%%%%%%%%%%%%
\textit{Conclusion--}
We investigated the coherent 2D spectrum of an interacting quantum many-body system of spin-1/2 fermions moving in a lattice, for example, electronic and ultracold atomic systems. 
In the weakly interacting limit, the 2D rephasing spectrum contains only diagonal peaks because different optical transitions are not coupled to one another.
In contrast, if the interaction between particles is sufficiently strong, off-diagonal peaks emerge, which can be used as a direct probe of a non-negligible interaction.
This is attributed to the fact that different optical transitions can be coupled to one another in the presence of interparticle interaction. 
Notably, effective coupling is induced by the quantum many-body interaction between particles as opposed to the conventional coupling between two transitions at the level of single-body physics.
As the interaction strength is increased further, when the system approaches the Mott insulating phase in the strongly interacting limit, all the peaks coalesce into a single peak at the origin of the 2D spectrum, i.e., at the transition frequency between the ground and excited bands.
Moreover, when the interaction is time-dependent, the information of its time variation can be obtained from the evolution of the coherent 2D spectrum as a function of the time delay between the second and third pulses.
The results of this study demonstrate the potential of coherent multidimensional spectroscopy for studying quantum many-body interaction and ultrafast dynamics in various kinds of strongly correlated systems.

%%%%%%%%%%%%%%%%%%%%%%%%%%
\begin{acknowledgements}
N. T. P. would like to thank Akihito Ishizaki and Yuta Fujihashi for discussion in preparing the manuscript.
This work was supported by JSPS KAKENHI Grant Numbers 19K14638 (N.~T.~Phuc).
The computations were performed using Research Center for Computational Science, Okazaki, Japan.
\end{acknowledgements}

%%%%%%%%%%%%%%%%%%%%%%%%%%

%%%%%%%%%%%%%%%

\begin{thebibliography}{100}
\bibitem{Landau56}
L. D. Landau,
The theory of a Fermi liquid.
J. Exptl. Theoret. Phys. (U.S.S.R.) \textbf{30}, 1058--1064 (1956).

\bibitem{Tomonaga50}
S. Tomonaga,
Remarks on Bloch's method of sound waves applied to many-fermion problems.
Prog. Theor. Phys. \textbf{5}, 544--569 (1950).

\bibitem{Luttinger63}
J. M. Luttinger,
An exactly soluble model of a many-fermion system.
J. Math. Phys. \textbf{4}, 1154 (1963).

\bibitem{Stewart01}
G. R. Stewart,
Non-Fermi-liquid behavior in d- and f-electron metals.
Rev. Mod. Phys. \textbf{73}, 797 (2001).

\bibitem{Mott68}
N. F. Mott,
Metal-insulator transition.
Rev. Mod. Phys. \textbf{40}, 677 (1968).

\bibitem{Imada98}
M. Imada, A. Fujimori, and Y. Tokura,
Metal-insulator transitions.
Rev. Mod. Phys. \textbf{70}, 1039 (1998).

\bibitem{Leggett-book}
A. J. Leggett,
\textit{Quantum Liquids: Bose condensation and Cooper pairing in condensed-matter systems} (Oxford University Press, New York, 2006).

%\bibitem{Bloch08}I. Bloch, J. Dalibard, and W. Zwerger,Many-body physics with ultracold gases.Rev. Mod. Phys. \textbf{80}, 885 (2008).

%\bibitem{Bloch12}I. Bloch, J. Dalibard, and S. Nascimbene,Quantum simulations with ultracold quantum gases.Nat. Phys. \textbf{8}, 267 (2012).

%\bibitem{Gross17}C. Gross and I. Bloch,Quantum simulations with ultracold atoms in optical lattices.Science \textbf{357}, 995--1001 (2017).

%\bibitem{Moses17}S. A. Moses, J. P. Covey, M. T. Miecnikowski, D. S. Jin, and J. Ye,New frontiers for quantum gases of polar molecules.Nat. Phys. \textbf{13}, 13 (2017).

%\bibitem{Bohn17}J. L. Bohn, A. M. Rey, and J. Ye,Cold molecules: Progress in quantum engineering of chemistry and quantum matter.Science \textbf{357}, 1002--1010 (2017).

%\bibitem{Campbell06}G. K. Campbell et al., Imaging the Mott Insulator Shells by Using Atomic Clock Shifts.Science \textbf{313}, 649 (2006).

%\bibitem{Kato16}S. Kato et al., Laser spectroscopic probing of coexisting superfluid and insulating states of an atomic Bose-Hubbard system.Nat. Commun. \textbf{7}, 11341 (2016).

\bibitem{Krausz09}
F. Krausz and M. Ivanov,
Attosecond physics.
Rev. Mod. Phys. \textbf{81}, 163 (2009).

\bibitem{Fausti11}
D. Fausti et al.,
Light-Induced Superconductivity in a Stripe-Ordered Cuprate.
Science \textbf{331}, 189 (2011).

\bibitem{Cavalleri18}
A. Cavalleri, 
Photo-induced superconductivity.
Contemporary Physics \textbf{59}, 31--46 (2018). 

\bibitem{Kimel05}
A. V. Kimel et al.,
Ultrafast non-thermal control of magnetization by instantaneous photomagnetic pulses.
Nature \textbf{435}, 655 (2005).

\bibitem{Kirilyuk10}
A. Kirilyuk, A. V. Kimel, and T. Rasing,
Ultrafast optical manipulation of magnetic order.
Rev. Mod. Phys. \textbf{82}, 2731 (2010).

\bibitem{Wang13}
Y. H. Wang, H. Steinberg, P. Jarillo-Herrero, and N. Gedik,
Observation of Floquet-Bloch States on the Surface of a Topological Insulator.
Science \textbf{342}, 453 (2013).

\bibitem{McIver20}
J. W. McIver et al., 
Light-induced anomalous Hall effect in graphene.
Nat. Phys. \textbf{16}, 38--41 (2020).

\bibitem{Mukamel-book}
S. Mukamel,
\textit{Principles of nonlinear optical spectroscopy}
(Oxford University Press, New York, 1995).

\bibitem{Mukamel00}
S. Mukamel,
Multidimensional femtosecond correlation spectroscopies of electronic and vibrational excitations.
Annu. Rev. Phys. Chem.\textbf{51}, 691--729 (2000).

\bibitem{Jonas03}
D. M. Jonas,
Two-dimensional femtosecond spectroscopy.
Annu. Rev. Phys. Chem. \textbf{54}, 425--463 (2003).

\bibitem{Cho08}
M. Cho,
Coherent Two-Dimensional Optical Spectroscopy.
Chem. Rev. \textbf{108}, 1331--1418 (2008).

\bibitem{Cho-book}
M. Cho,
\textit{Two-dimensional optical spectroscopy}
(CRC Press, Taylor \& Francis Group, 2009).

\bibitem{Abramavicius09}
D. Abramavicius, B. Palmieri, D. V. Voronine, F. Sanda, and S. Mukamel,
Coherent Multidimensional Optical Spectroscopy of Excitons in Molecular Aggregates; Quasiparticle versus Supermolecule Perspectives.
Chem. Rev. \textbf{109}, 2350--2408 (2009).

\bibitem{Hamm-book}
P. Hamm and M. Zanni,
\textit{Concepts and methods of 2D infrared spectroscopy}
(Cambridge University Press, New York, 2011).

\bibitem{Cundiff09}
S. T. Cundiff, T. Zhang, A. D. Bristow, D. Karaiskaj, and X. Dai,
Optical Two-Dimensional Fourier Transform Spectroscopy of Semiconductor Quantum Wells.
Acc. Chem. Res. \textbf{42}, 1423--1432 (2009).

\bibitem{Cohen11}
G. S. Schlau-Cohen, A. Ishizaki, and G. R. Fleming,
Two-dimensional electronic spectroscopy and photosynthesis: Fundamentals and applications to photosynthetic light-harvesting.
Chem. Phys. \textbf{386}, 1--22 (2011).

\bibitem{Kuehn11}
W. Kuehn, K. Reimann, M. Woerner, T. Elsaesser, and R. Hey,
Two-Dimensional Terahertz Correlation Spectra of Electronic Excitations in Semiconductor Quantum Wells.
J. Phys. Chem. B \textbf{115}, 5448--5455 (2011).

\bibitem{Woerner13}
M. Woerner, W. Kuehn, P. Bowlan, K. Reimann, and T. Elsaesser,
Ultrafast two-dimensional terahertz spectroscopy of elementary excitations in solids.
New J. Phys. \textbf{15}, 025039 (2013).

\bibitem{Lu17}
J. Lu et al., 
Coherent Two-Dimensional Terahertz Magnetic Resonance Spectroscopy of Collective Spin Waves.
Phys. Rev. Lett. \textbf{118}, 207204 (2017).

\bibitem{Wan19}
Y. Wan and N. P. Armitage,
Resolving Continua of Fractional Excitations by Spinon Echo in THz 2D Coherent Spectroscopy.
Phys. Rev. Lett. \textbf{122}, 257401 (2019).

\bibitem{Choi20}
W. Choi, K. H. Lee, and Y. B. Kim,
Theory of Two-Dimensional Nonlinear Spectroscopy for the Kitaev Spin Liquid.
Phys. Rev. Lett. \textbf{124}, 117205 (2020).

\bibitem{Mahmood20}
F. Mahmood, D. Chaudhuri, S. Gopalakrishnan, R. Nandkishore, and N. P. Armitage,
Observation of a marginal Fermi glass using THz 2D coherent spectroscopy.
arXiv:2005.10822 (2020).

\bibitem{Novelli20}
F. Novelli, J. O. Tollerud, D. Prabhakaran, and J. A. Davis,
Persistent coherence of quantum superpositions in an optimally doped cuprate revealed by 2D spectroscopy.
Sci. Adv. \textbf{6}, eaaw9932 (2020).

%\bibitem{Dai10}X. Dai, A. D. Bristow, D. Karaiskaj, and S. T. Cundiff,Two-dimensional Fourier-transform spectroscopy of potassium vapor.Phys. Rev. A \textbf{82}, 052503 (2010).

%\bibitem{Khalil03}M. Khalil, N. Demirdoven, and A. Tokmakoff,Obtaining Absorptive Line Shapes in Two-Dimensional Infrared Vibrational Correlation Spectra.Phys. Rev. Lett. \textbf{90}, 047401 (2003).

%\bibitem{Zheng05}J. Zheng, K. Kwak, J. Asbury, X. Chen, I. R. Piletic, and M. D. Fayer,Ultrafast Dynamics of Solute-Solvent Complexation Observed at Thermal Equilibrium in Real Time.Science \textbf{309}, 1338 (2005).

%\bibitem{Gessner14}M. Gessner, F. Schlawin, H. Haffner, S. Mukamel, and A. Buchleitner,Nonlinear spectroscopy of controllable many-body quantum systems.New J. Phys. \textbf{16}, 092001 (2014).

%\bibitem{Schlawin14}F. Schlawin, M. Gessner, S. Mukamel, and A. Buchleitner,Nonlinear spectroscopy of trapped ions.Phys. Rev. A \textbf{90}, 023603 (2014).

\bibitem{Gullans12}
M. Gullans et al., 
Nanoplasmonic lattices for ultracold atoms.
Phys. Rev.  Lett. \textbf{109}, 235309 (2012).

\bibitem{Tudela15}
A. Gonzalez-Tudela, C.-L. Hung, D. E. Chang, J. I. Cirac and H. J. Kimble,
Subwavelength vacuum lattices and atom–atom interactions in two-dimensional photonic crystals.
Nat. Photonics \textbf{9}, 320--325 (2015).

\bibitem{Nascimbene15}
S. Nascimbene, N. Goldman, N. R. Cooper, and J. Dalibard,
Dynamic Optical Lattices of Subwavelength Spacing for Ultracold Atoms.
Phys. Rev. Lett. \textbf{115}, 140401 (2015).

\bibitem{Isart13}
O. Romero-Isart, C. Navau, A. Sanchez, P. Zoller, and J. I. Cirac,
Superconducting Vortex Lattices for Ultracold Atoms.
Phys. Rev. Lett. \textbf{111}, 145304 (2013).


\end{thebibliography}
\end{document}